\documentclass[conference]{IEEEtran}
\IEEEoverridecommandlockouts
\usepackage{cite}
\usepackage{amsmath,amssymb,amsfonts}
\usepackage{algorithmic}
\usepackage{graphicx}
\usepackage{textcomp}
\usepackage{xcolor}
\usepackage{cite}
\usepackage{amsmath,amssymb,amsfonts}
\usepackage{algorithmic}
\usepackage{fancybox}
\usepackage{pdfpages}
\usepackage{comment}
\usepackage{graphicx}
\usepackage{mathtools, cuted}
\usepackage{lipsum, color}
\usepackage{comment}
\DeclareUnicodeCharacter{00A0}{~}
\usepackage{textcomp}
\usepackage{xcolor}
\usepackage{cite}
\usepackage{amsmath,amssymb,amsfonts}
\usepackage{algorithmic}
\usepackage{graphicx}
\usepackage{textcomp}

\def\BibTeX{{\rm B\kern-.05em{\sc i\kern-.025em b}\kern-.08em
    T\kern-.1667em\lower.7ex\hbox{E}\kern-.125emX}}
\begin{document}

\title{\textit{Invited:} Can Wi-R enable perpetual IoB nodes?\\
\thanks{\textcolor{black}{This work was supported by the National Science Foundation Career Award under Grant CCSS 1944602}}

{\Large \textit{}}\vspace{-0.5em}
}

\author{\IEEEauthorblockN{Arunashish Datta, \textit{Student Member, IEEE} and Shreyas Sen, \textit{Senior Member, IEEE}}
\IEEEauthorblockA{\textit{Elmore Family School of Electrical and Computer Engineering, Purdue University, West Lafayette, USA}}
\IEEEauthorblockA{\textit{E-mail: datta30@purdue.edu, shreyas@purdue.edu}}
}
\maketitle

\begin{abstract}
While the number of wearables is steadily growing, the wearables/person wearing them faces a limitation due to the need for charging all of them every day. To unlock the true power of IoB, we need to make these IoB nodes perpetual. However, that is not possible with today's technology. In this paper, we will debate, whether with the advent of Wi-R protocol that uses the body to communicate at $100X$ lower energy that BTLE$/$Wi-Fi, is it going to be possible to enable the long-standing desire of perpetual sensing$/$actuation nodes for the Internet of Bodies.
\end{abstract}


\section{Introduction}

\begin{figure}[t]
\centering
\includegraphics[width=0.5\textwidth]{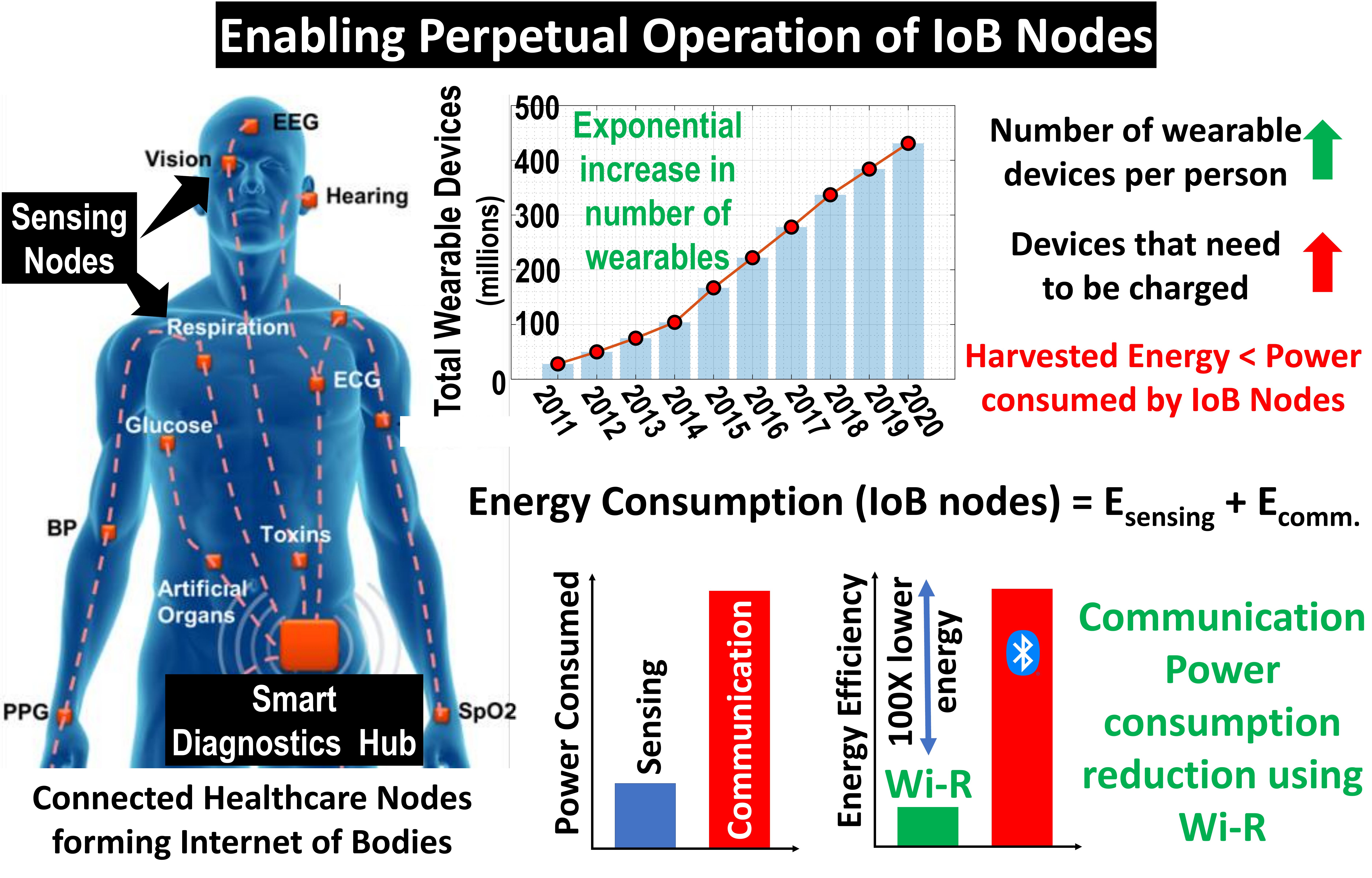}
\caption{Increasing wearables/person, has resulted in a need for techniques to enable perpetual operation for the nodes to reduce time spent in charging devices. Wi-R has come up as an exciting alternative to traditional RF based techniques, demonstrating higher communication energy efficiency.}
\label{fig:intro}
\vspace{-1.5em}
\end{figure}

Wearable devices are increasing exponentially over the last decade. This interconnected network of devices around the body forms the subsection of Internet of Things (IoT) termed as Internet of Bodies or IoB (Fig. \ref{fig:intro}). IoB \cite{cIoB,TEDX,maity2017secure,Rand_IoB_Connected_Future,Rand,Rand_Better_or_Worse,Forbes_IoB,Chatterjee_Bioelectronic} is fast becoming a reality with the continuous scaling of devices resulting in increasing number of devices in and around the body. However, for an efficient implementation of IoB, it is critical to analyze the bottlenecks currently preventing its widespread adoption and the potential solutions. \par
The most important application for IoB nodes is in the continuous health monitoring by sensing biopotential signals from the human body. Smartphones and smartwatches tracking and providing updates on daily health has become commonplace. However, more accurate and denser sensing technologies require higher number of IoB nodes across the body. Considerable research is being done to develop more biopotential signal measurement devices. However, increasing number of wearable devices per person means charging more and more devices and further keeping track of these devices through the day. This is a serious threat to the adoption of IoB as charging these devices can become an extremely tedious task. Thus, perpetual operation of IoB nodes has become a topic of extensive research. 
\par
The vision of perpetual IoB nodes is being pursued in the research community in a multipronged approach. Wearable IoB nodes are size constrained and thus have small batteries. Extensive research is being done to store higher energy in confined spaces to increase energy density in batteries. However, battery technology scaling has not caught up to the scaling of semiconductor technology. Researchers have investigated reducing the power of sensing and communication for IoB nodes, as well looking at new methods for energy harvesting for charging. Radio frequency (RF) based wireless powering methodologies have been looked into with commercial applications developed over the last few years. RF based wireless powering however has still got a long way to go before they enable perpetual operation of current IoB nodes. This is because of the difficulty in transmitting power over the air, with channel losses frequently above $60 dB$ due to fading and shadowing from objects in the vicinity. Further, the transmit power is also limited by safety concerns, which are guided by the Federal Communications Commissions (FCC).   \par
Therefore, the logical conclusion is to design lower power devices that can run for longer time with the current wireless powering and harvesting systems. It is well known that the power consumption in the IoB nodes is dominated by the communication power, which is typically $3-4$ orders of magnitude above the computation power consumed \cite{sen2016context, Chatterjee_Bioelectronic}. Thus, a major research thrust has been in the development of innovative communication methodologies tailor-made for operation of devices in and around the body. Typically, data is communicated between IoB nodes using radio frequency (RF) based radiative communication protocols, which transmit signals over the air. Common examples of such communication protocols include Bluetooth, ZigBee and LoRa. However, these communication methodologies have been shown to be insecure due to their radiative nature, and have high power consumption, a large part of which is lost to the signal that is radiated across the room. This decreases the communication energy efficiency by increasing the amount of power required to successfully communicate each bit. \par
Thus, in the pursuit of energy efficient communication, directed methods of communicating between two nodes placed around the body have been explored so that power is not lost in radiating data across the air channel. This has prompted research into using the body as a wire to communicate data between IoB nodes, as the human body serves as a common medium connecting all the devices. The use of body as a wire to communicate data was efficiently exploited by using Electro-Quasistatic Human Body Communication (EQS-HBC) \cite{maity2018bio, datta2021advanced, modak2020bio, Safety_Study, datta2021channel}. EQS-HBC has been demonstrated to be orders of magnitude more energy efficient than typical RF based protocols and has been shown to be physically secure by confining the signal within the body, making it unavailable to attackers unless they are almost in contact with the body. \\
Wi-R technology is the first commercial implementation of EQS-HBC. Wi-R has orders of magnitude higher energy efficiency when compared to traditional RF based methods of communication. Wi-R technology operates at $1-20 MHz$ frequency range which falls firmly within the realms of Electro-Quasistatic frequencies for Human Body Communication. In this study, we present a method for unified analysis of sensing and communication systems to investigate the use of Wi-R in place of RF based communication strategies and the doors that this will open for the perpetual operation of IoB nodes.     
\begin{figure}[b]
\centering
\includegraphics[width=0.5\textwidth]{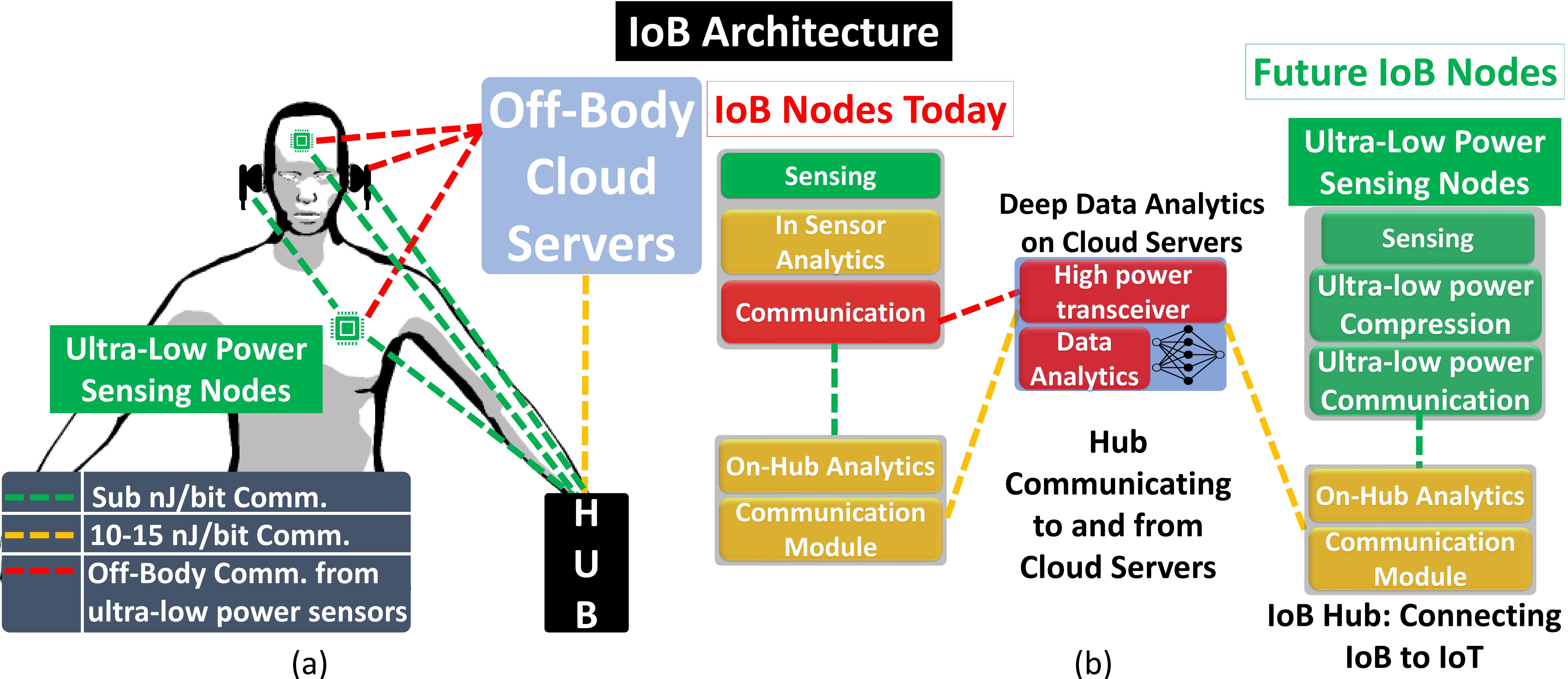}
\caption{Architecture and functions of low-power sensing nodes and hubs in IoB. (a) IoB network architecture. (b) Current sensing structure and possible future low-power sensing nodes.}
\label{fig:IoB_Arch}
\end{figure}
\section{IoB Devices and the need for Wi-R}

IoB devices can be broadly categorized into two different types based on their functions and form factor: (i) Larger hubs, and (ii) Smaller nodes. The smaller nodes are the various sensors in and around the body, which may be fitness trackers, connected healthcare devices, implanted or ingested medical devices, physiological sensors or a combination of all of them. These devices are ultra-low power nodes required to run without having to be charged for a long period of time, ideally perpetually. The larger devices are the hubs, which bear most of the computation demands and are responsible for connecting IoB to the larger IoT for further data analytics on the cloud. These are devices like Smartphones and Smartwatches, AR/VR smart glasses and headsets and consume higher power than the smaller IoB nodes.\par
An architecture of the proposed IoB system is illustrated in Fig. \ref{fig:IoB_Arch}. The smaller nodes cannot communicate data at all times to and from cloud servers off the body, as illustrated in Fig. \ref{fig:IoB_Arch} (a). To design ultra-low power small IoB nodes, a bulk of the communication and computation needs to be offloaded to the hubs. A comparison between current sensor node designs and future ultra-low power IoB sensing nodes is shown in Fig. \ref{fig:IoB_Arch} (b). Thus, the main tasks of these nodes can be divided into three parts: (i) Sensing, (ii) Ultra-low power compression, and (iii) Communicating sensed data to the hub. This ensures that the higher power demands in data analytics and communicating data off the body are offloaded to the larger hubs. Ultra-low power compression specially for audio and video signals is key in reducing data volume. However, in this study, we focus on the benchmarking of sensing and communication systems without delving into the details of compression. \\
The energy requirements for the ultra-low power IoB nodes can be mathematically denoted as illustrated by Eqn: \ref{eqn:E_total},
\begin{equation}
    E_{total} = E_{sensing} + E_{comm.}
    \label{eqn:E_total}
\end{equation}
where, the total energy consumption $(E_{total})$ is the sum of the energy requirement for sensing $(E_{sensing})$ and the communication energy required $(E_{comm.})$. This can be further broken down to analyze the dependence of $E_{total}$ on the energy efficiency or the energy spent per bit of sensing and communication  $(\eta_{sensing}, \eta_{comm.})$, as shown by Eqn. \ref{eqn:eff}.
\begin{equation*}
    E_{sensing} = bit_{sensed} \times \frac{E_{sensing}}{bit}
\end{equation*}
\begin{equation*}
    E_{sensing} = bit_{sensed} \times \eta_{sensing}
\end{equation*}
\begin{equation*}
    E_{comm.} = bit_{sensed} \times \frac{E_{comm.}}{bit}
\end{equation*}
\begin{equation*}
    E_{comm.} = bit_{sensed} \times \eta_{comm.}
\end{equation*}
\begin{equation}
    E_{total} = bit_{sensed} \times (\eta_{sensing} + \eta_{comm.})
    \label{eqn:eff}
\end{equation}
Thus, it is essential to increase energy efficiency or decrease $\eta$ of sensing and communication to ensure that the smaller nodes have a high lifetime before they are charged. In this study, we investigate energy efficient methods of communication between IoB nodes. \\
Bluetooth has been the gold standard for communication between devices around the body in commercial devices. However, in recent times, EQS-HBC has been demonstrated to be an energy efficient and physically secure alternative to traditional RF based communication techniques specifically for IoB devices. In EQS-HBC, data is transmitted through the body using the tissue's conductive properties to communicate information between two devices on the body. The signal is coupled to the body at frequencies $< 40 MHz$. At these frequencies, the human body is an inefficient antenna and the data is confined within the body. Thus, no power is lost to data being transmitted through the air. This ensures that EQS-HBC is physically secure and has a higher energy efficiency than RF based radiative communication techniques. \par
Wi-R is the first commercial application of the EQS-HBC technology \cite{senWiR}. Wi-R uses non-radiative near field communication using EQS fields across the human body. Wi-R has shown immense potential in bringing the EQS-HBC technology proposed in literature to the masses. Wi-R has orders of magnitude higher energy efficiency when compared to Bluetooth, which makes it the prime candidate for around the body communication in IoB nodes. Current Wi-R technology also allows for comparable or higher data rates around the body at a fraction of the power spent when compared to Bluetooth. This can potentially be a game changer for smaller IoB nodes communicating to the hub by bringing us a step closer to perpetual operation of these ultra-low power nodes. We now analyze the sensing and communication power requirements as well as the available power to the nodes, either via energy harvesting from wireless powering methodologies or traditionally using smaller batteries.
\begin{figure}[t]
\centering
\includegraphics[width=0.5\textwidth]{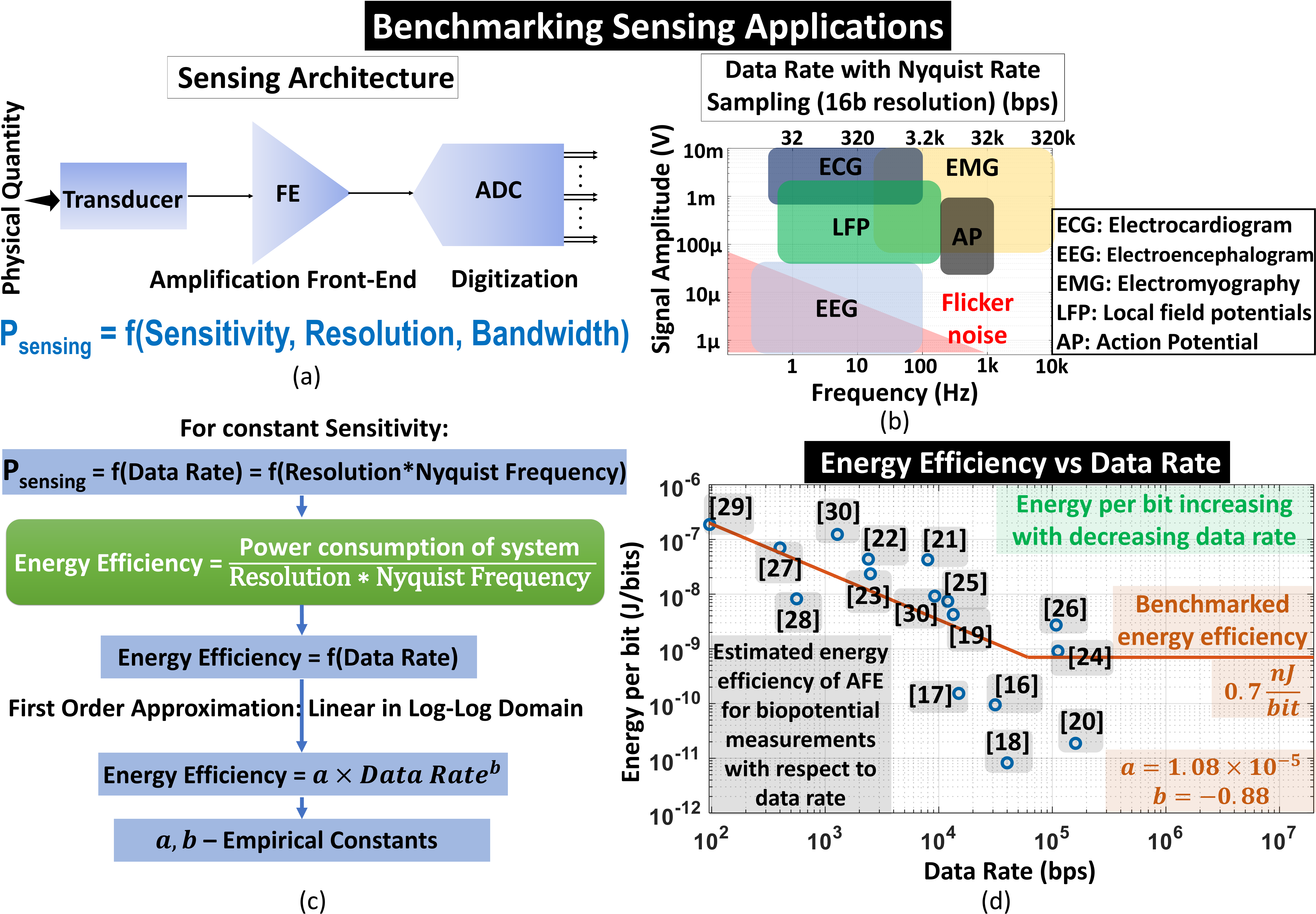}
\caption{Benchmarking $\eta_{sensing}$. (a) A typical sensing node. (b) Data rates and sensitivities for common biopotential signals. The algorithm in (c) is used to simplify and benchmark $\eta_{sensing}$. (d) Survey of literature and commercial space and benchmarked energy efficiency.}
\label{fig:sensing}
\vspace{-1em}
\end{figure}

\section{Benchmarking different energies}
 A benchmarking of sensing and communication power as well as amount of harvested energy is performed to estimate the performance of ultra-low power IoB nodes operating with Wi-R and Bluetooth. The absolute values of energy-efficiency for sensing and communication systems vary vastly with changing architectures, and is not essential to this study. We perform a study to simplify and approximate the energy-efficiency as a function of data rate, considering no variance in other parameters. This simplified view aids in developing a unified analysis of sensing and communication techniques.  
\subsection{Sensing}
A typical sensing architecture is illustrated in Fig. \ref{fig:sensing} (a). For such sensing nodes, the power consumption is a function of the sensitivity, resolution and the bandwidth. IoB nodes typically deal with sensing biopotential signals, some of which are illustrated in Fig. \ref{fig:sensing} (b). Fig. \ref{fig:sensing} (b) also shows the required sensitivity for these applications and the minimum data rate required to capture these signals with a $16-bit$ ADC. 
To benchmark sensing power as a function of data rate, we perform a survey of studies in the literature and for commercial analog front ends for biopotential signal acquisition \cite{8946728,6964818,7231342,6637111,7599667,8662407,6757498,van2014345,rieger2018integrated,o2018recursive,schonle2018multi,9365757,8662404,8846577,MAX30001}. The energy efficiency for the whole system is estimated based on the algorithm present in Fig. \ref{fig:sensing} (c). The vast variability in sensing power due to different applications and circuit architectures leads us to attempt to simplify the analysis. The algorithm is a first order approximation to estimate the variation in energy efficiency as a function of data rate for iso-sensitivity and iso-design analog front ends. The collected data for energy efficiency as a function of data rate is fitted to a line in the log-log domain using MATLAB, to estimate the energy efficiency of such a system. The curve plateaus at $0.7nJ/bit$ to approximate the energy efficiency for higher data rates. This curve illustrated in Fig. \ref{fig:sensing} (d), is used as a benchmark for energy efficiency of sensing systems $(\eta _ {sensing})$ for future analysis of ultra-low power IoB sensing nodes. 
\begin{figure}[b]
\centering
\includegraphics[width=0.5\textwidth]{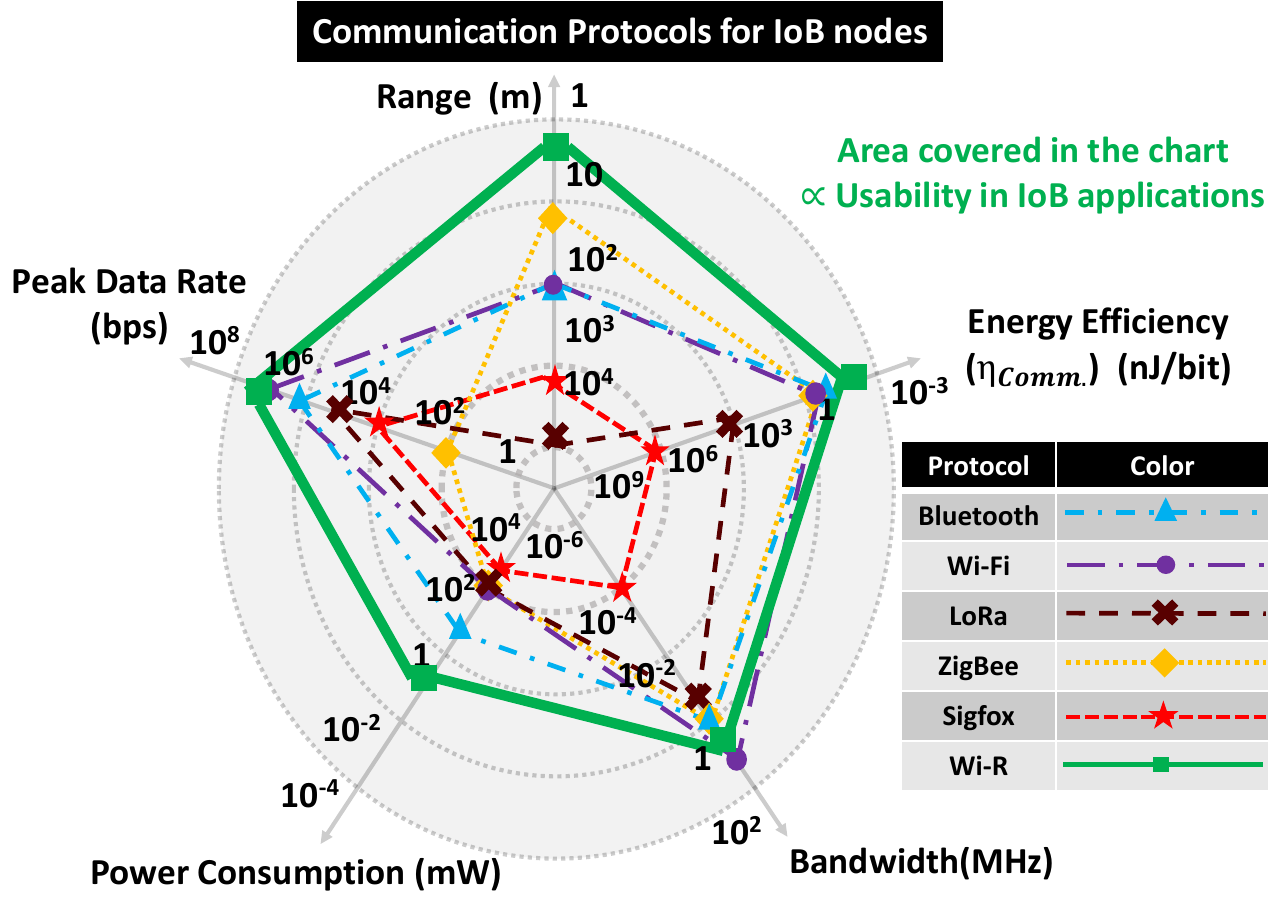}
\caption{Communication protocols for IoB nodes.}
\label{fig:Communication_IoB}
\end{figure}
\begin{figure*}[t]
\centering
\includegraphics[width=0.9\textwidth]{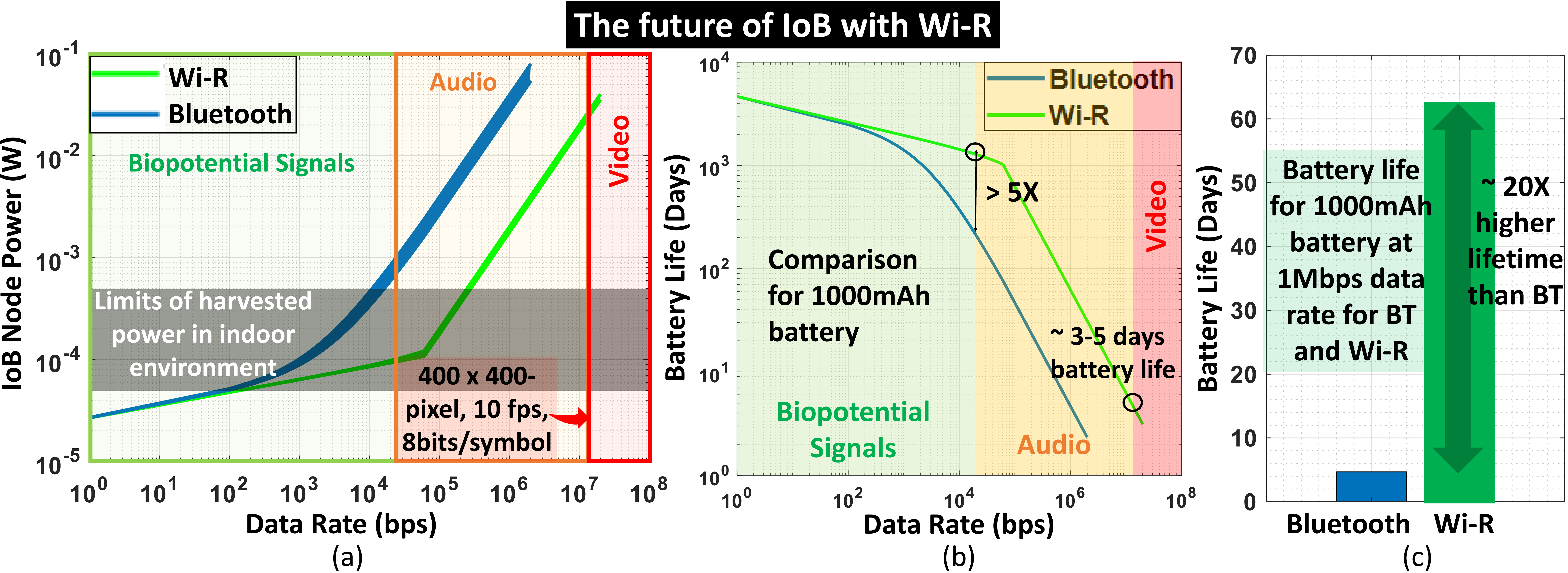}
\vspace{-0.5em}
\caption{(a) Comparison between power consumption for IoB sensor nodes operating with Bluetooth (blue) and Wi-R (green) as a function of data rate. (b) Device lifetime as a function of data rate for a $1000 mAh$ battery. (c) Comparison of battery life for a $1000 mAh$ battery for a data rate of $1Mbps$ with Bluetooth (red) and Wi-R (green).}
\label{fig:results}
\vspace{-1.5em}
\end{figure*}
\vspace{-0.5em}
\subsection{Communication}
Communication technologies for IoB nodes require energy efficient communication for channel lengths of $1cm-2m$. Benchmarking current communication technologies requires us to look at the commonly used communication techniques for IoB nodes which are typically RF based protocols like Bluetooth, LoRa and ZigBee. However, these RF based communication protocols can consume $> 1nJ/bit$ energy efficiency and up to $15nJ/bit$ for Bluetooth. \par
Emerging technology in the form of EQS-HBC has been shown to be much more energy efficient \cite{JSSC,maity2021sub}. Wi-R, the first commercial implementation of EQS-HBC has $<100pJ/bit$ energy efficiency for communication around the body \cite{senWiR}. Future iterations of Wi-R are aimed at bridging the gap between literature and consumer applications to bring the energy efficiency closer to $10pJ/bit$. A concept diagram describing the various different communication protocols and their key parameters (Energy Efficiency, Data rate, range etc.) is illustrated in Fig. \ref{fig:Communication_IoB}, where the higher is the area occupied by a particular protocol, the better it is for use in IoB nodes. Wi-R being specifically designed to operate around the body can be seen to have immense potential in IoB applications.  
\vspace{-0.2em}
\subsection{Available Energy for Wearables}
Wearable and implantable ultra-low power IoB devices being size constrained require using small sized batteries and further depend on wireless powering and energy harvesting mechanisms to operate. Typically, the batteries used are small coin cell batteries ($200mAh$ \cite{CR2032} to about $1000 mAh$ \cite{CR2477}), as these are safe to use and can be operated for long durations.\par
To further extend the lifetime of ultra-low power IoB sensor nodes, various different methods for harvesting energy for IoB nodes has come up over the last decade. These can be broadly categorized into mechanical, thermal, radiative, chemical, magnetic and electric. A study of energy harvesting modalities and the peak power density for the different modalities has been conducted by Chatterjee et al. \cite{Chatterjee_Bioelectronic}. In an indoor setup with variability due to position of devices and postures, realistically, the amount of harvested power is between $50 \mu W$ to $100s$ of $\mu W$ \cite{avlani2022eico, 8050615}. \par

\section{Impact of Wi-R on IoB Nodes}
The impact of Wi-R on IoB nodes is illustrated using Fig. \ref{fig:results}. Fig. \ref{fig:results} (a) shows the estimated power of an ultra-low power IoB node as a function of data rate using the benchmarked energy efficiencies for communication and sensing. The data and the source code used to generate the graphs are provided publicly \cite{Github_graph}. We consider a system efficiency factor: $\eta_{system} = 0.4$ which emulates the additional losses coming from LDOs, DC-DC converters and the platform power consumption. The area occupied by the blue and green curves in Fig. \ref{fig:results} (a) illustrate the variation in power consumption of commercial versions and studies in literature. The knee points on the blue and green curves come from the plateauing of sensing energy efficiency at $0.7 nJ/bit$. Harvested power is limited to between $50 \mu W$ to $100s$ of $\mu W$ for indoor environments, as shown in the figure. We observe that with state-of-the-art harvesting techniques, Bluetooth allows reliable perpetual operation for low data rate ($\approx 1 bps - 100s bps$) biopotential sensing applications. Continuous biopotential signal monitoring with higher data rates is unlikely using Bluetooth. In comparison, Wi-R can be used to operate ultra-low power biopotential sensors perpetually for almost the whole spectrum of biopotential signals. Further, current Wi-R technology with an energy efficiency of $< 100 pJ/bit$ has the potential to transmit low data rate audio signals ($\leq 200 kbps$) reliably and operate perpetually with state-of-the-art harvesting techniques. 
Fig. \ref{fig:results} (b) and (c) shows the improvement in device lifetime for a battery operated ($1000mAh$) node when using Wi-R as compared to Bluetooth. Using Wi-R can increase battery life by more than an order of magnitude for audio and low data rate video ($400 \times 400$ pixels, $10fps$, $8$ bits/symbol) applications without compression. Incorporating ultra-low power compression techniques can further enhance the reach of Wi-R to higher data rate applications, increasing the spectrum of applications possible.  \par
The current sensing systems with Bluetooth can operate perpetually only for low data rate applications with current energy harvesting techniques. We have demonstrated that implementing Wi-R enables perpetual operation for most biopotential sensing applications and improves battery life by more than an order of magnitude for audio/video applications. 
\vspace{-0.5em}
\section{Conclusion}
We present a framework for a unified analysis of sensing, and communication methodologies to study the existing state-of-the-art for communication around the body and compare with Wi-R, an exciting new alternative which uses the body as a directed wire for communication between devices around the body. Wi-R is $100X$ more energy efficient $(<100pJ/bit)$ as compared to traditional Bluetooth $(>15nJ/bit)$ based communication platforms, and opens up the potential for a future where more devices can operate perpetually.

\bibliographystyle{IEEEtran}

\bibliography{references}

\begin{thebibliography}{10}
\providecommand{\url}[1]{#1}
\csname url@samestyle\endcsname
\providecommand{\newblock}{\relax}
\providecommand{\bibinfo}[2]{#2}
\providecommand{\BIBentrySTDinterwordspacing}{\spaceskip=0pt\relax}
\providecommand{\BIBentryALTinterwordstretchfactor}{4}
\providecommand{\BIBentryALTinterwordspacing}{\spaceskip=\fontdimen2\font plus
\BIBentryALTinterwordstretchfactor\fontdimen3\font minus
  \fontdimen4\font\relax}
\providecommand{\BIBforeignlanguage}[2]{{%
\expandafter\ifx\csname l@#1\endcsname\relax
\typeout{** WARNING: IEEEtran.bst: No hyphenation pattern has been}%
\typeout{** loaded for the language `#1'. Using the pattern for}%
\typeout{** the default language instead.}%
\else
\language=\csname l@#1\endcsname
\fi
#2}}
\providecommand{\BIBdecl}{\relax}
\BIBdecl

\bibitem{cIoB}
\BIBentryALTinterwordspacing
{Purdue University}. (2019) {Center for Internet of Bodies}. [Online].
  Available: \url{https://engineering.purdue.edu/C-IoB}
\BIBentrySTDinterwordspacing

\bibitem{TEDX}
\BIBentryALTinterwordspacing
{Shreyas Sen}. (2019) {How your body will play an integral role in the future
  of wearable security}. [Online]. Available:
  \url{https://www.ted.com/talks/shreyas\_sen\_how\_your\_body
  \_will\_play\_an\_integral\_role\_in\_the\_future\_of\_wearable \_security}
\BIBentrySTDinterwordspacing

\bibitem{maity2017secure}
S.~Maity, D.~Das, X.~Jiang, and S.~Sen, ``Secure human-internet using dynamic
  human body communication,'' in \emph{2017 IEEE/ACM International Symposium on
  Low Power Electronics and Design (ISLPED)}.\hskip 1em plus 0.5em minus
  0.4em\relax IEEE, 2017, pp. 1--6.

\bibitem{Rand_IoB_Connected_Future}
\BIBentryALTinterwordspacing
{Rand Corporation}. (2022) {Internet of Bodies: Our Connected Future}.
  [Online]. Available:
  \url{https://www.rand.org/about/nextgen/art-plus-data/giorgia-lupi/internet-of-bodies-our-connected-future.html}
\BIBentrySTDinterwordspacing

\bibitem{Rand}
M.~Lee, B.~Boudreaux, R.~Chaturvedi, S.~Romanosky, and B.~Downing, \emph{The
  internet of bodies: Opportunities, risks, and governance}.\hskip 1em plus
  0.5em minus 0.4em\relax Rand Corporation, 2020.

\bibitem{Rand_Better_or_Worse}
\BIBentryALTinterwordspacing
{Rand Corporation}. (2020) {Internet of Bodies: Our Connected Future}.
  [Online]. Available:
  \url{https://www.rand.org/about/nextgen/art-plus-data/giorgia-lupi/internet-of-bodies-our-connected-future.html}
\BIBentrySTDinterwordspacing

\bibitem{Forbes_IoB}
\BIBentryALTinterwordspacing
{Forbes}. (2019) {What Is The Internet Of Bodies? And How Is It Changing Our
  World?} [Online]. Available:
  \url{https://www.forbes.com/sites/bernardmarr/2019/12/06/what-is-the-internet-of-bodies-and-how-is-it-changing-our-world/?sh=269b998968b7}
\BIBentrySTDinterwordspacing

\bibitem{Chatterjee_Bioelectronic}
\BIBentryALTinterwordspacing
B.~Chatterjee, P.~Mohseni, and S.~Sen, ``Bioelectronic sensor nodes for the
  internet of bodies,'' \emph{Annual Review of Biomedical Engineering},
  vol.~25, no.~1, p. null, 2023, pMID: 36913705. [Online]. Available:
  \url{https://doi.org/10.1146/annurev-bioeng-110220-112448}
\BIBentrySTDinterwordspacing

\bibitem{sen2016context}
S.~Sen, ``Context-aware energy-efficient communication for iot sensor nodes,''
  in \emph{Proceedings of the 53rd Annual Design Automation Conference}, 2016,
  pp. 1--6.

\bibitem{maity2018bio}
S.~Maity, M.~He, M.~Nath, D.~Das, B.~Chatterjee, and S.~Sen, ``Bio-physical
  modeling, characterization, and optimization of electro-quasistatic human
  body communication,'' \emph{IEEE Transactions on Biomedical Engineering},
  vol.~66, no.~6, pp. 1791--1802, 2018.

\bibitem{datta2021advanced}
A.~Datta, M.~Nath, D.~Yang, and S.~Sen, ``Advanced biophysical model to capture
  channel variability for {EQS} capacitive {HBC},'' \emph{IEEE Transactions on
  Biomedical Engineering}, 2021.

\bibitem{modak2020bio}
N.~Modak \emph{et~al.}, ``Bio-physical modeling of galvanic human body
  communication in electro-quasistatic regime,'' \emph{IEEE Transactions on
  Biomedical Engineering}, vol.~69, no.~12, pp. 3717--3727, 2022.

\bibitem{Safety_Study}
S.~{Maity et al.}, ``On the safety of human body communication,'' \emph{IEEE
  Transactions on Biomedical Engineering}, pp. 1--1, 2020.

\bibitem{datta2021channel}
A.~Datta, M.~Nath, B.~Chatterjee, N.~Modak, and S.~Sen, ``Channel modeling for
  physically secure electro-quasistatic in-body to out-of-body communication
  with galvanic tx and multimodal rx,'' in \emph{2021 IEEE MTT-S International
  Microwave Symposium (IMS)}.\hskip 1em plus 0.5em minus 0.4em\relax IEEE,
  2021, pp. 116--119.

\bibitem{senWiR}
S.~Sen, ``Wi-r technology white paper,'' \emph{Ixana}, 2023.

\bibitem{8946728}
J.~P. Uehlin, W.~A. Smith, V.~R. Pamula, S.~I. Perlmutter, J.~C. Rudell, and
  V.~S. Sathe, ``A 0.0023 mm$^2$/ch. delta-encoded, time-division multiplexed
  mixed-signal ecog recording architecture with stimulus artifact
  suppression,'' \emph{IEEE Transactions on Biomedical Circuits and Systems},
  vol.~14, no.~2, pp. 319--331, 2020.

\bibitem{6964818}
R.~Muller, H.-P. Le, W.~Li, P.~Ledochowitsch, S.~Gambini, T.~Bjorninen,
  A.~Koralek, J.~M. Carmena, M.~M. Maharbiz, E.~Alon, and J.~M. Rabaey, ``A
  minimally invasive 64-channel wireless $\mu$ecog implant,'' \emph{IEEE
  Journal of Solid-State Circuits}, vol.~50, no.~1, pp. 344--359, 2015.

\bibitem{7231342}
A.~E. Mendrela, J.~Cho, J.~A. Fredenburg, C.~A. Chestek, M.~P. Flynn, and
  E.~Yoon, ``Enabling closed-loop neural interface: A bi-directional interface
  circuit with stimulation artifact cancellation and cross-channel cm noise
  suppression,'' in \emph{2015 Symposium on VLSI Circuits (VLSI Circuits)},
  2015, pp. C108--C109.

\bibitem{6637111}
W.-M. Chen, H.~Chiueh, T.-J. Chen, C.-L. Ho, C.~Jeng, M.-D. Ker, C.-Y. Lin,
  Y.-C. Huang, C.-W. Chou, T.-Y. Fan, M.-S. Cheng, Y.-L. Hsin, S.-F. Liang,
  Y.-L. Wang, F.-Z. Shaw, Y.-H. Huang, C.-H. Yang, and C.-Y. Wu, ``A fully
  integrated 8-channel closed-loop neural-prosthetic cmos soc for real-time
  epileptic seizure control,'' \emph{IEEE Journal of Solid-State Circuits},
  vol.~49, no.~1, pp. 232--247, 2014.

\bibitem{7599667}
B.~C. Raducanu, R.~F. Yazicioglu, C.~M. Lopez, M.~Ballini, J.~Putzeys, S.~Wang,
  A.~Andrei, M.~Welkenhuysen, N.~van Helleputte, S.~Musa, R.~Puers,
  F.~Kloosterman, C.~van Hoof, and S.~Mitra, ``Time multiplexed active neural
  probe with 678 parallel recording sites,'' in \emph{2016 46th European
  Solid-State Device Research Conference (ESSDERC)}, 2016, pp. 385--388.

\bibitem{8662407}
J.~Warchall, P.~Theilmann, Y.~Ouyang, H.~Garudadri, and P.~P. Mercier, ``22.2 a
  rugged wearable modular exg platform employing a distributed scalable
  multi-channel fm-adc achieving 101db input dynamic range and motion-artifact
  resilience,'' in \emph{2019 IEEE International Solid- State Circuits
  Conference - (ISSCC)}, 2019, pp. 362--364.

\bibitem{6757498}
J.~Xu, B.~Busze, H.~Kim, K.~Makinwa, C.~Van~Hoof, and R.~F. Yazicioglu, ``24.7
  a 60nv/$\sqrt{Hz}$ 15-channel digital active electrode system for portable
  biopotential signal acquisition,'' in \emph{2014 IEEE International
  Solid-State Circuits Conference Digest of Technical Papers (ISSCC)}, 2014,
  pp. 424--425.

\bibitem{van2014345}
N.~Van~Helleputte, M.~Konijnenburg, J.~Pettine, D.-W. Jee, H.~Kim, A.~Morgado,
  R.~Van~Wegberg, T.~Torfs, R.~Mohan, A.~Breeschoten \emph{et~al.}, ``A 345
  $\mu$w multi-sensor biomedical soc with bio-impedance, 3-channel ecg, motion
  artifact reduction, and integrated dsp,'' \emph{IEEE Journal of Solid-State
  Circuits}, vol.~50, no.~1, pp. 230--244, 2014.

\bibitem{rieger2018integrated}
R.~Rieger and M.~Rif’an, ``Integrated exg, vibration and temperature
  measurement front-end for wearable sensing,'' \emph{IEEE Transactions on
  Circuits and Systems I: Regular Papers}, vol.~65, no.~8, pp. 2422--2430,
  2018.

\bibitem{o2018recursive}
G.~O'Leary, M.~R. Pazhouhandeh, M.~Chang, D.~Groppe, T.~A. Valiante, N.~Verma,
  and R.~Genov, ``A recursive-memory brain-state classifier with 32-channel
  track-and-zoom $\delta$ 2 $\sigma$ adcs and charge-balanced programmable
  waveform neurostimulators,'' in \emph{2018 IEEE International Solid-State
  Circuits Conference-(ISSCC)}.\hskip 1em plus 0.5em minus 0.4em\relax IEEE,
  2018, pp. 296--298.

\bibitem{schonle2018multi}
P.~Sch{\"o}nle, F.~Glaser, T.~Burger, G.~Rovere, L.~Benini, and Q.~Huang, ``A
  multi-sensor and parallel processing soc for miniaturized medical
  instrumentation,'' \emph{IEEE Journal of Solid-State Circuits}, vol.~53,
  no.~7, pp. 2076--2087, 2018.

\bibitem{9365757}
Q.~Lin, S.~Song, R.~V. Wegberg, M.~Konijnenburg, D.~Biswas, C.~V. Hoof,
  F.~Tavernier, and N.~V. Helleputte, ``28.3 a 28$\mu$w 134db dr 2nd-order
  noise-shaping slope light-to-digital converter for chest ppg monitoring,'' in
  \emph{2021 IEEE International Solid- State Circuits Conference (ISSCC)},
  vol.~64, 2021, pp. 390--392.

\bibitem{8662404}
A.~Caizzone, A.~Boukhayma, and C.~Enz, ``17.8 a 2.6$\mu$w monolithic cmos
  photoplethysmographic sensor operating with 2$\mu$w led power,'' in
  \emph{2019 IEEE International Solid- State Circuits Conference - (ISSCC)},
  2019, pp. 290--291.

\bibitem{8846577}
H.~Ha, W.~Sijbers, R.~Van~Wegberg, J.~Xu, M.~Konijnenburg, P.~Vis,
  A.~Breeschoten, S.~Song, C.~Van~Hoof, and N.~V. Helleputte, ``A bio-impedance
  readout ic with digital-assisted baseline cancellation for two-electrode
  measurement,'' \emph{IEEE Journal of Solid-State Circuits}, vol.~54, no.~11,
  pp. 2969--2979, 2019.

\bibitem{MAX30001}
\BIBentryALTinterwordspacing
{Analog Devices}. {MAX30001, Ultra-Low-Power, Single-Channel Integrated
  Biopotential (ECG, R-to-R, and Pace Detection) and Bioimpedance (BioZ) AFE}.
  [Online]. Available: \url{https://www.analog.com/en/products/max30001.html}
\BIBentrySTDinterwordspacing

\bibitem{JSSC}
S.~{Maity} \emph{et~al.}, ``Bodywire: A 6.3-pj/b 30-mb/s -30-db {SIR}-tolerant
  broadband interference-robust human body communication transceiver using time
  domain interference rejection,'' \emph{IEEE Journal of Solid-State Circuits},
  vol.~54, no.~10, pp. 2892--2906, Oct 2019.

\bibitem{maity2021sub}
S.~Maity, N.~Modak, D.~Yang, M.~Nath, S.~Avlani, D.~Das, J.~Danial,
  P.~Mehrotra, and S.~Sen, ``Sub-$\mu$wrcomm: 415-nw 1-10-kb/s physically and
  mathematically secure electro-quasi-static hbc node for authentication and
  medical applications,'' \emph{IEEE Journal of Solid-State Circuits}, 2021.

\bibitem{CR2032}
\BIBentryALTinterwordspacing
{Energizer Holdings, Inc.} {Energizer CR2032, Produce Datasheet}. [Online].
  Available: \url{https://data.energizer.com/pdfs/cr2032.pdf}
\BIBentrySTDinterwordspacing

\bibitem{CR2477}
\BIBentryALTinterwordspacing
{Panasonic Energy}. {CR2477 CR Coin-type Lithium Battery}. [Online]. Available:
  \url{https://industrial.panasonic.com/cdbs/www-data/pdf2/AAA4000/AAA4000C341.pdf}
\BIBentrySTDinterwordspacing

\bibitem{avlani2022eico}
S.~Avlani, D.-H. Seo, B.~Chatterjee, and S.~Sen, ``Eico: Energy-harvesting
  long-range environmental sensor nodes with energy-information dynamic
  co-optimization,'' \emph{IEEE Internet of Things Journal}, vol.~9, no.~21,
  pp. 20\,932--20\,944, 2022.

\bibitem{8050615}
P.~Jokic and M.~Magno, ``Powering smart wearable systems with flexible solar
  energy harvesting,'' in \emph{2017 IEEE International Symposium on Circuits
  and Systems (ISCAS)}, 2017, pp. 1--4.

\bibitem{Github_graph}
\BIBentryALTinterwordspacing
{SparcLab}. {Perpetual IoB, Wi-R and BLE}. [Online]. Available:
  \url{{https://github.com/SparcLab/Perpetual-IoB\_Wi-R-BLE.git}}
\BIBentrySTDinterwordspacing

\end{thebibliography}

\end{document}